\documentclass[11pt]{article}
\usepackage{epsfig}
\usepackage{epsf}

\def \ksks {K^*\bar K^*}
\def \jp {J/\psi}
\def \ra {\rangle}
\def \la {\langle}
\def \fc {F_{\textrm{color}}}

\def \fss {F_{\textrm{spin+space}}}

\def \half {{1\over 2}}
\newcommand {\cg}[3]{\langle #1;#2|#3\rangle}
\newcommand {\sla}[1]{ #1 \!\!\!/}
\newcommand{\epon}{\epsilon_1}
\newcommand{\eptw}{\epsilon_2}

\title{Study of $X(1812)$ as a $(K^*\bar K^*)$ Molecular State}
\author{Hong Chen$^a$, Rong-Gang Ping$^b$\\
a) School of Physical Science and Technology, Southwest
University,\\
Chongqing 400715, People's Republic of China. \\
b) Institute of High Energy Physics,Chinese Academy of Sciences,\\
 P.O. Box 918(1), Beijing 100049, China;\footnote{Mailing address}\footnote{E-mail:pingrg@mail.ihep.ac.cn}\\
\date{ }}

\begin{document}
\maketitle

\abstract{We investigate the decay and the production mechanism of
the resonance $X(1812)$ recently observed in the $\jp\to\gamma
X(1812),~X(1812)\to\omega\phi$ at BESII. The decay widths of
$X(1812)\to\eta\eta'$,$~\eta\eta$,$~\omega\phi$,$~K^+K^-$,$~\rho^+\rho^-$,
$~\omega\omega$,$~K^{*+}K^{*-}$ and $\pi^+\pi^-$ are evaluated based
on the scenario of the $X(1812)$ as a candidate of $(\ksks)$
molecule. It turns out that the quark exchange mechanism plays an
important role in the understanding of the large decay width for the
$X(1812)\to\omega\phi$. It is also found that the decay widths for
$X(1812)\to\eta\eta'$ and $\eta\eta$ are enhanced by the quark
exchange mechanism. These channels are suggested to be the tools to
test the molecular scenario in experiment. The branching fraction of
$Br(X\to\omega\phi)$ is evaluated to be about $4.60\%$. Searches for
additional evidence about the $X(1812)$ in $\jp$ radiative decays
are reviewed. In the molecular scenario, the $X(1812)$ production
rate is also evaluated to be $\Gamma(\jp\to\gamma
X)/\Gamma(\jp\to\gamma K^{*+}K^{*-})=2.13^{+7.41}_{-1.85}$, which is
close to the measured
value $2.83\pm0.92$.}\\
\vspace{0.0cm}
 \vspace{1.0cm}
\section{Introduction}
The observation of a near-threshold enhancement in $\jp\to\gamma
X,~X\to\omega\phi$ at BESII \cite{bes_x} immediately provokes
discussion about its nature. This enhancement is reported to favor
$J^{PC}=0^{++}$ in a partial wave analysis with a mass and width of
$M=1812^{+19}_{-26}(\textrm{stat})\pm18(\textrm{syst})$ and
$\Gamma=105\pm 20(\textrm{stat})\pm28(\textrm{syst})$ MeV,
respectively, and a production branching ratio, $B(\jp\to\gamma
X)\cdot
B(X\to\omega\phi)=[2.61\pm0.27(\textrm{stat})\pm0.65(\textrm{syst})]\times
10^{-4}$. This resonant state is named as $X(1812)$ in the following
part.

Emergence of the $X(1812)$ adds a new puzzle to the scalar sector of
mesons. The most interesting feature about $X(1812)$  is that it
seems to have a strange production and decay mechanism observed in
the $\jp\to\gamma\omega\phi$. If the valence quarks of $\omega$ and
$\phi$ are assigned as $(u\bar u+d\bar d)/\sqrt 2$ and $s\bar s$,
one expects that this double OZI suppressed process should have a
smaller branching fraction, at least less than the OZI allowed
processes, e.g. $\jp\to\gamma\phi\phi$ or $\gamma\omega\omega$. It
seems that the $X(1812)$ has a large contribution to the
$\jp\to\gamma\omega\phi$ decay. Moreover, if the scalar of $X(1812)$
is assigned as the $q\bar q$ ordinary meson, it weakly couples to
the decay mode of $\omega\phi$ and has a small phase space to decay
into $\omega\phi$ near the threshold. However, the observed
branching fraction is at the order of $10^{-4}$. So it is difficult
to fit the $X(1812)$ into the spectrum of ordinary $q\bar q$ mesons.
Many efforts have been made to interpret the $X(1812)$ as an exotic
state\cite{ex1,ex2,ex3}, such as glueball\cite{gl1,gl2}, hybrid
\cite{hy1,hy2} and four-quark state \cite{fq}. The contributions
from final state interactions of $f_0(1710)$ decays into $PP$ and
$VV$ states are already investigated \cite{fs}.

Search for ordinary decay modes of the $X(1812)$ is essential for
understanding its structure. Like the other scalar mesons, e.g.
$f_0(1500)$, one expects that the $X(1812)$ may have a large
fraction to decay into $PP$ and $VV$ states. However, no more
experimental information on $X(1812)$ are available, a thorough
investigation on all possible production and decay mechanisms for
$X(1812)$ is thus necessary for understanding data, such as the
molecule scenario should be inspected. The wide resonance $K^{*}$
and its pair $(\ksks)$ near the $X(1812)$ threshold make it a
candidate for such consideration. PDG \cite{pdg06} quotes
$M=891.66\pm0.26$ MeV and $\Gamma=50.8\pm0.9$ MeV as the average
value of the mass and total width for the $K^{*\pm}$, and almost the
same values for the $K^{*0}$. If the width of a resonance is
regarded as the full width at half maximum of the mass distribution
in a nonrelativistic Breit-Wigner form, the $\ksks$ pair still has a
large probability to lay within the $X(1812)$ mass region. Moreover,
the production of $\ksks$ is copious in the $\jp\to\gamma\ksks$ with
the branching fraction
$Br(\jp\to\gamma\ksks)=(4.0\pm1.3)\times10^{-3}$ \cite{pdg06}. Some
interactions between the $K^*$ and $\bar K^*$ can not be ruled out,
thus give rise to the reaction $\ksks\to\omega\phi$ possibly.

The theoretical studies on the scalar meson structure are always the
controversial subjects in particle physics, and the sparse
experimental information makes it more difficult to draw a decisive
conclusion about their nature. Even for the well established scalar
$f_0(980)$ and $a_0(980)$, there are still more puzzles about their
masses, decay rates and so on. Many theoretical efforts have been
made in literatures to study the scalar as the exotic meson, such as
the four-quark state \cite{jaffe}, glueball or hybrid \cite{glu} and
molecular state \cite{mol}. In the quark potential model, the
molecule of $\ksks$ is predicted to be $f_0(1710)$ \cite{chenjx}
with a bound energy about $70$ MeV. However, the quark model based
on the pairwise effective interactions predicted that there exists a
weak bound molecule near the threshold, such as the $D\bar D$
molecule \cite{wongcy} with the bound energy about 3 MeV. Near the
$\ksks$ threshold, the possibility that some interactions between
$\ksks$ pair might give rise to a weakly bound molecular state can
not be ruled out.

In brief, the production and decay mechanism of the $X(1812)$
observed in the $\jp\to\gamma\omega\phi$ deserve to do a thorough
investigation, especially by assigning the $X(1812)$ as a molecular
state, from which we expect to learn more about $X(1812)$. The
$(\ksks)$ pair could be a candidate of such consideration. As
follows, we will formulate the $X(1812)$ decay and production rates
by assuming the $X(1812)$ as the $\ksks$ molecule, from which we
hope to see whether the experimental data could be understood in
this scenario, and which channel could be possibly used to look for
the $X(1812)$ in experiment and to find its trace again. Therefore,
all possible descriptions about the $X(1812)$ structure are subject
to the experimental test in the future, and the comparison between
the data and the theoretical prediction may shed some light on the
$X(1812)$ nature.
\section{Decay Mechanism}
The experimental information on the X(1812) shows that the
$J^{PC}=0^{++}$ is favored. So, in the scenario of $(\ksks)$
molecular state, the $\ksks$ is chosen as the orbital ground state
$L=0$ and the spin singlet $S=0$ due to the restriction of the
space-orbit parity $[P=(-1)^L]$ and the charge parity
$[C=(-1)^{L+S}]$, the wave function is expressed as \cite{chenjx}:
\begin{equation}\label{mole_wave}
\Psi_{(\ksks)}=\sum_{m_1,m_2}C^{00}_{1m_1,1m_2}\epsilon(m_1)\epsilon(m_2)\varphi_{K^*}({\bf
p_1})\varphi_{\bar K^*}({\bf p_2})\Psi_{(\ksks)}({\bf p_1-p_2}),
\end{equation}
where $\epsilon(m_i)$ is the spin wave function of $K^*$ or $\bar
K^*$, and the $\varphi_K(p_i)$ is the $K^*$ or $\bar K^*$ wave
function in momentum space.
\begin{figure}[htbp]

\begin{center}
\hspace*{-.cm} \epsfysize=3cm \epsffile{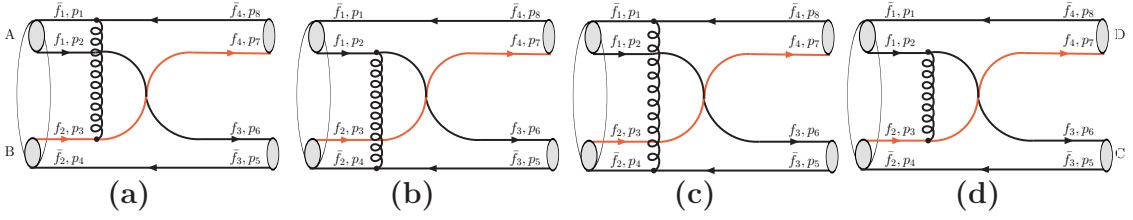}
\parbox{0.8\textwidth}{\caption{The four meson-meson scattering diagrams by exchanging
$f_1f_2$ quarks, and similarly by exchanging $\bar f_1\bar f_1$
quarks. \label{exch}}}
\end{center}
\end{figure}

\begin{figure}[htbp]
 \vspace*{0cm}
\begin{center}
\hspace*{-.cm} \epsfysize=3cm \epsffile{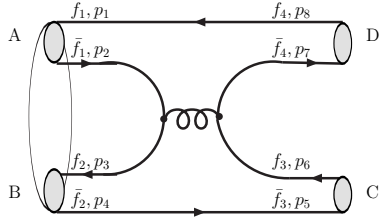} \caption{The
schematic diagram for decays of the $\ksks$ molecule via a $q\bar q$
annihilation.\label{anni}}
\end{center}
\end{figure}
At hadronic level, the $\ksks$ can be scattered into the $\omega
\phi$ final sate by exchanging a meson  as assumed in \cite{fs}. At
quark level, such a meson-meson scatterring process has already been
investigated by many authors in quark model \cite{qqs1}-\cite{qqs5}.
It is found that the process can be described by one gluon exchange
(OGE) between $qq$ or $\bar q\bar q$ pair, by which one found that
it gives an excellent description both for the light and heavy
meson-meson scattering process. By analogy with these models, we
describe the reaction $\ksks\to MM$ to begin with the quark-quark
scattering, then leads to the quark rearrangement between the two
color singlet clusters and subsequent formation of the final two
mesons as shown in Fig. \ref{exch}(a-d). In addition, the $q\bar q$
annihilation may take place between $K^*$ and $\bar K^*$ clusters
and leads to the $(\ksks)$ molecule to decay into the different
final state as shown in Fig. \ref{anni}.

 As the $\ksks$ in the
molecule are weakly bound, we may relate this decay rate to the
cross section $\sigma(\ksks\to M_1M_2)$ near threshold, by analogy
with the calculation of the decay rate of $^1S_0$ positronium to
$\gamma\gamma$. The decay rate for a $(\ksks)$-molecule is
\cite{cross_scatter}:
\begin{equation}
\Gamma ((\ksks)\to M_1M_2)=\sigma(\ksks\to M_1M_2)\cdot v_{rel}\cdot
|\Psi_{(\ksks)}(0)|^2,
\end{equation}
where $v_{rel}$ is the relative velocity of the $K^*$ and $\bar
K^*$, $\Psi_{(\ksks)}(0)$ is the $(\ksks)$-molecule wavefunction at
the origin ${\bf r}=0$. The cross section can be expressed by the
invariant amplitudes for the scattering of $\ksks$ to the final
meson pair, i.e.
\begin{equation}\label{cross}
{d\sigma((\ksks)\to M_1M_2)\over dt}={1\over 64\pi s^2}{1\over
|P_{Acm}|^2}|\mathcal M((\ksks)\to M_1M_2)|^2,
\end{equation}
where the Mandlestam variable $t=M_A^2-2E_AE_C+2{\bf P_A\cdot
P_C}+M_C^2$.
 The invariant amplitude $\mathcal M$ is generally expressed as:
\begin{equation}\label{}
\mathcal{M}=\la
M_1M_2|H_I|(\ksks)\ra=F_{\textrm{color}}F_{\textrm{flavor}}F_{\textrm{spin+space}}
\end{equation}
For the quark-exchange mechanism and the quark-annihilation
mechanism, on has $\fc=4/9$.

The flavor factors are calculated based on the flavor functions of
the final meson pairs with the interaction $H_f$, which can be
written as, for example,
\begin{equation}
H_f=\left \{
\begin{array}{lr}
\delta_{f_1f_3}\delta_{f_2f_4}\delta_{\bar f_1\bar f_4}\delta_{\bar
f_2\bar f_3}& \textrm{ for Fig. \ref{exch} (a)},\\
a^+_{\bar f_4}a^+_{f_3}a_{\bar f_1}a_{f_2} & \textrm{ for Fig.
\ref{anni} }.
\end{array}\right.
\end{equation}
where $f$ denotes quark flavor, $a^+_f$ and $a_f$ are the quark $f$
creation and annihilation operator, respectively.

The contributions from the spin and space parts are denoted by:
\begin{equation}
\fss=\la\varphi_{M_1}\varphi_{M_2}|H_{ss}|\varphi_{K^*}\varphi_{\bar
K^*}\ra,
\end{equation}
where $\varphi_{M}$ and $\varphi_{K^*}$ denote the wave functions of
the decayed mesons and $K^*$ molecular state in spin and coordinate
space, respectively. They are  directly evaluated by Feynman
diagrams as shown in Fig. \ref{exch} and  Fig. \ref{anni}. It reads:
\begin{eqnarray}
\fss&=&{1\over (2\pi)^6}\int \left[\prod_{i=1}^8 d^3{\bf p}_i\right
] \mathcal{H}~\varphi_{K^*}({\bf p_1-p_2})\varphi_{\bar K^*}({\bf
p_3-p_4})\varphi_{\bar
M_1}({\bf p_5-p_6})\varphi_{M_2}({\bf p_7-p_8})\nonumber\\
&\times&\delta^3({\bf P_A-p_1-p_2 })\delta^3({\bf P_B-p_3-p_4
})\delta^3({\bf P_C-p_5-p_6 })\delta^3({\bf P_D-p_7-p_8 }),
\end{eqnarray}

For the quark $f_1f_2$ exchange mode (Fig. \ref{exch}), the operator
$\mathcal H$ reads:
\begin{eqnarray}
\mathcal H_{ex}^{(a)}={1\over (p_7-p_3)^2}\bar v(s_1,p_1)\gamma_\mu
v(s_8,p_8)\bar u(s_7,p_7)\gamma^\mu
u(s_3,p_3)\delta_{s_4,s_5}\delta_{s_2,s_6}\delta^3({\bf p_4-p_5})\delta^3({\bf p_2-p_6}),\\
\mathcal H_{ex}^{(b)}={1\over (p_6-p_3)^2}\bar v(s_4,p_4)\gamma_\mu
v(s_5,p_5)\bar u(s_6,p_6)\gamma^\mu
u(s_2,p_2)\delta_{s_2,s_7}\delta_{s_1,s_8}\delta^3({\bf
p_2-p_7})\delta^3({\bf
p_1-p_8}),\\
\mathcal H_{ex}^{(c)}={1\over (p_5-p_4)^2}\bar v(s_1,p_1)\gamma_\mu
v(s_8,p_8)\bar v(s_4,p_4)\gamma^\mu
v(s_5,p_5)\delta_{s_2,s_6}\delta_{s_3,s_7}\delta^3({\bf
p_2-p_6})\delta^3({\bf
p_3-p_7}),\\
\mathcal H_{ex}^{(d)}={1\over (p_6-p_2)^2}\bar u(s_6,p_6)\gamma_\mu
u(s_2,p_2)\bar u(s_7,p_7)\gamma^\mu
u(s_3,p_3)\delta_{s_1,s_8}\delta_{s_4,s_5}\delta^3({\bf
p_1-p_8})\delta^3({\bf p_4-p_5}),
\end{eqnarray}
where $u(s_i,p_i)$ and $v(s_i,p_i)$ are the quark and antiquark
Dirac spinor with the normalization condition $u^\dag(s,p)
u(s,p)=-v^\dag(s,p) v(s,p)=2p^0$. For the quark $\bar f_1\bar f_2$
exchange mode, it has the same diagram and the operator $\mathcal
H_{ex}$ takes the similar form.

For the quark $\bar f_1 f_2$ and $f_1\bar f_2$ annihilation mode
(Fig. \ref{anni}), the interaction $\mathcal H$ are respectively
given by:
\begin{eqnarray}
\mathcal H_{an}^{\bar f_1f_2}={1\over (p_2+p_3)^2}\bar
v(s_7,p_7)\gamma_\mu u(s_6,p_6)\bar u(s_3,p_3)\gamma^\mu
v(s_2,p_2)\delta_{s_1,s_8}\delta_{s_4,s_5}\delta^3({\bf
p_1-p_8})\delta^3({\bf
p_4-p_5}),\\
\mathcal H_{an}^{f_1\bar f_2}={1\over (p_1+p_4)^2}\bar
u(s_1,p_1)\gamma_\mu v(s_4,p_4)\bar v(s_5,p_5)\gamma^\mu
u(s_8,p_8)\delta_{s_2,s_7}\delta_{s_3,s_6}\delta^3({\bf
p_2-p_7})\delta^3({\bf p_3-p_6}),
\end{eqnarray}

\section{Production Mechanism}
The production of the molecule $(\ksks)$ is assumed via the $\jp$
radiative decay into a real photon ($\gamma$) plus two virtual
gluons ($gg$), followed by the formation of the meson pair of
$\ksks$, and then possible formation of the $\ksks$ molecule through
the final state interactions occurs. To note that the decay
$\jp\to\gamma \ksks$ has a large branching fraction (Br=$(4.0\pm
1.3)\times 10^{-3})$, then the interactions between the $\ksks$ pair
may take place and may give birth to the $(\ksks)$ molecule.
\begin{figure}[htbp]
 \vspace*{0cm}
\begin{center}
\hspace*{-.cm} \epsfysize=3cm \epsffile{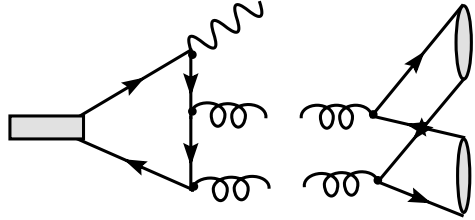}\\
\parbox{0.6\textwidth}{ \caption{The schematic diagram for the
production of the $(\ksks)$ molecule in $\jp$ radiative
decays.\label{figprd}}}
\end{center}
\end{figure}

To evaluate the production rate of $(\ksks)$-molecule, we consider
the process as shown in Fig. \ref{figprd} at the level of the
leading order of perturbative QCD, and the bound states are
phenomenologically described by their wavefunctions. The whole
amplitude can be decomposed into two parts as done in
\cite{korner,amp_deco}, which is written as
\begin{equation}\label{amp}
A=\sum {1\over k_1^2k_2^2}\langle \jp|\gamma g g\rangle\langle g
g|(\ksks)\rangle,
\end{equation}
where $k_1$ and $k_2$ are the momentum of the two virtual off-shell
gluons. The matrix elements of $\langle \jp|\gamma g g\rangle$ and
$\langle g g|(\ksks)\rangle$ describes the subprocess $\jp\to \gamma
gg$ and $gg\to (K^*\bar K^*)$, respectively. The sum is over the
polarization vectors of the two gluons.

For the subprocess $\jp(E)\to \gamma(\epsilon,k)
g(\epon,k_1)g(\eptw,k_2)$, the amplitude can be expressed by
\cite{korner}:
\begin{equation}\label{part1}
A^{\psi\to\gamma
gg}_{\alpha\nu\nu_1\nu_2}E^\alpha\epsilon^{*\nu}\epsilon^{*\nu_1}_1\epsilon^{*\nu_2}_2=
8i{R_{\psi}(0)\over \sqrt{4\pi M^3}}{M^2a_v\over (k_1+k_2)\cdot
k(k+k_1)\cdot k_2(k+k_2)\cdot k_1},
\end{equation}
with
\begin{eqnarray}
a_v&=&\{\epon^*\cdot \eptw^*[-k_1\cdot k\epsilon\cdot k_2E\cdot
k_1-k_2\cdot k\epsilon^*\cdot k_1E\cdot k_2-k_1\cdot kk_2\cdot
kE\cdot \epsilon^*]\nonumber \\
&+&E\cdot \epsilon^*[k_1\cdot k\epon^*\cdot k_2\eptw\cdot k+k_2\cdot
k\eptw^*\cdot k_1\epon^*\cdot
k-k_1\cdot k_2\epon^*\cdot k\eptw^*\cdot k]\}\nonumber \\
&+&\{\epon,k_1\leftrightarrow
\epsilon,k\}+\{\eptw,k_2\leftrightarrow \epsilon,k\},
\end{eqnarray}
where $E, \epon$ and $\eptw$ are polarization vectors for $\jp$, the
real photon and the two virtual gluons, respectively, and their
momentum are denoted by $k, k_1$ and $k_2$, respectively.
$R_{\psi}(0)$ is the $\jp$ radial wave function at origin in
coordinate space.

For the subprocess $g(\epon,k_1)g(\eptw,k_2)\to
K^*(\epsilon_{K^*},P_2)\bar K^*(\epsilon_{\bar K^*},P_2)$, the
amplitude can be directly calculated from the leading-order diagram
as shown in Fig. \ref{figprd} by using the standard Feymman rules.
The details of the calculation are given in appendix A. The
amplitude is calculated to be:
\begin{eqnarray}\label{part2}
 A_{gg\to K^*\bar K^*}^{\mu_1,\mu_2}\epsilon_{1\mu_1}\epsilon_{2\mu_2}&=&{1\over 2}{1\over 4\sqrt{p_1^0p_2^0q_1^0q_2^0}}
 {1\over
 \sqrt{(q_1^0+m_s)(q_2^0+m_s)(p_1^0+m_u)(p_2^0+m_u)}}{R_{K^*}^2(0)\over 4\pi}\nonumber \\
&\times& \textrm{Tr}[(-\sla q_1+m_s)\sla \epsilon_{K^*}(\sla
p_1+m_u)\sla\epsilon_2(-\sla q_2+m_s)\epsilon_{\bar K^*}(\sla
p_2+m_u)\sla \epsilon_1\nonumber \\
&+&(\epon\leftrightarrow \eptw)],
\end{eqnarray}
where $R_{K^*}(0)$ is the $K^*$ wave function at origin, and $p_1$
and $q_1$ are respectively the momentum of $u-$quark and $s-$quark
bound in $K^*$, and the antiparticle counterparts denoted by $p_2$
and $q_2$ with $p_i=m_uP_i/(m_s+m_u)$ and $q_i=m_sP_i/(m_s+m_u)$.

If we assume the $K^*\bar K^*$ pair to be a molecular state, which
is phenomenologically described by a wavefunction $\Phi(\bf P_1-
P_2)$ given in Eq. (\ref{mole_wave}), then the amplitude for
$gg\to(\ksks)$ reads:
\begin{equation}\label{}
 A_{gg\to (K^*\bar
 K^*)}^{\mu_1,\mu_2}\epsilon_{1\mu_1}\epsilon_{2\mu_2}=\int {d^3{\bf q}\over \sqrt{2q^0(2\pi)^3}}A^{gg\to K^*\bar
 K^*}_{\mu_1,\mu_2}\epsilon_{1\mu_1}\epsilon_{2\mu_2} {\Psi_{(\ksks)}(\bf q)},
\end{equation}

After inserting Eq. (\ref{part1}) and (\ref{part2}) into Eq.
(\ref{amp}), and making substitution $k_1=p_1+q_2$ and
$k_2=p_2+q_1$, one obtains the amplitude for $\jp\to\gamma \ksks$.
Using the standard formula for three-body decays \cite{pdg06}, the
decay width for $\jp\to \gamma K^*\bar K^*$ is written as:
\begin{equation}\label{prod1}
\Gamma={(2\pi)^4\over M_\psi}\sum^{-\!\!-}\int \left |
{g_{\nu_1\mu_1}g_{\nu_2\mu_2}\over (p_1+q_2)^2(p_2+q_1)^2}
A_{\psi\to\gamma gg}^{\alpha\nu\nu_1\nu_2}E_\alpha\epsilon^*_{\nu}
A_{gg\to K^*\bar K^*}^{\mu_1,\mu_2} \right |^2 d\phi_3,
\end{equation}
where $\phi_3$ is a standard 3-body phase space factor.

Similarly, the decay width for $\jp\to \gamma (K^*\bar K^*)$ is
written as:
\begin{equation}\label{prod2}
\Gamma={1 \over 8\pi}\sum^{-\!\!-}\left |
{g_{\nu_1\mu_1}g_{\nu_2\mu_2}\over (p_1+q_2)^2(p_2+q_1)^2}
A_{\psi\to\gamma gg}^{\alpha\nu\nu_1\nu_2}E_\alpha\epsilon^*_{\nu}
A_{gg\to (K^*\bar K^*)}^{\mu_1,\mu_2} \right |^2 {|{\bf
P_\gamma}|\over M_\psi^2},
\end{equation}
where ${\bf P_\gamma}$ is the photon momentum.
\section{Numerical Results}
To evaluate the $(\ksks)$ molecule decay width and the production
rate, one should have a reliable method to describe the
wavefunctions of the low lying mesons. However, no rigorous theory
from the QCD first principle is available to describe the light
bound states, so one has to use a phenomenological model to include
the nonperturbative properties. Phenomenologically, we construct
meson wave functions in the constituent quark model; they are
decomposed into three parts, i.e. the flavor, spin and the space
wavefunctions. For example, the wave function of $K^{*+}$ is
constructed as:
\begin{equation}
|K^{*+}\ra=|u\bar s\ra\epsilon(1,m)\varphi({\bf p}),
\end{equation}
where $\epsilon(1,m)$ denotes the symmetry spin wave function of two
constituent quarks. For pseudoscalar mesons, it is chosen as
asymmetric spin wave functions.

For $\eta$ and $\eta'$ mesons, we choose the mixing scheme to expand
their flavor functions in terms of the singlet and octet quark
flavor basis as:
\begin{eqnarray}
|\eta\ra&=&\cos\phi|\eta_q\ra-\sin\phi|\eta_s\ra, \nonumber\\
|\eta'\ra&=&\sin\phi|\eta_q\ra+\cos\phi|\eta_s\ra,
\end{eqnarray}
where $\phi$ is the mixing angle, $|\eta_q\ra=|u\bar u+d\bar
d\ra/\sqrt 2\varphi_q({\bf p})$ and $|\eta_s\ra=|s\bar
s\ra\varphi_s({\bf p})$, where $\varphi({\bf p})$ is chosen as the
ground harmonic oscillator basis associated with the decay constant
$f_q$ or $f_s$. In \cite{FKS} the mixing angle and the decay
constants are extracted from the experimental data as:
$\phi=39.3^\circ\pm 1.0^\circ,~f_q=(1.07\pm0.02)f_\pi$ and
$f_s=(1.34\pm0.06)f_\pi$. The flavor factors for
$X(1812)\to\eta\eta',~\eta\eta$, together with the other final
states, are given in Table \ref{flavorfactor}.

The space wave function $\varphi({\bf p})$ is solely dependent on
the harmonic oscillator parameter $\beta$. From some studies on the
the spectroscopy and decay rates of the low lying scalar and vector
mesons, their decay constant is determined; they are related to the
meson wavefunction at the origin by $f_{P,V}=\sqrt{12\over
M_{P,V}}|\phi_{P,V}(0)|$. On the other hand, the wave function at
the origin is related to the parameter $\beta$ by
$\phi_{P,V}(0)=(\beta^2/ \pi)^{3/4}$. Table \ref{harmonicpar}
summarizes the decay constants determined by experimental data
\cite{decay_constant} and the harmonic oscillator parameters
determined by the relation $\beta_{P,V}=\sqrt \pi ({M\over
12})^{1/3}f_{P,V}^{2/3}$.

\begin{table}
\begin{center}
\parbox{0.6\textwidth}{\caption{The flavor factor for the $\ksks$ molecule decays into a
meson pair via the quark exchange and annihilation
mechanism.\label{flavorfactor}}}
\begin{tabular}{cccc}
\hline\hline
        &\multicolumn{2}{c}{Annihilation}&Exchange\\\hline
$M_1M_2$&$s\bar s\to u\bar u$&$u\bar u\to s\bar s$&
$s\leftrightarrow u$ or $\bar s\leftrightarrow\bar u$\\\hline
$\omega\omega$&1/2&0&0\\
$\rho^+\rho^-$&1&0&0\\
$\omega\phi$&0&0&$1/\sqrt 2$\\
$K^{*+}K^{*-}$&1&1&0\\
$\eta\eta$&$\cos^2\phi/2$&$\sin^2\phi$&$-\sqrt 2\sin2\phi$\\
$\eta\eta'$&$\sin2\phi/4$&$-\sin^2\phi/2$&$\cos2\phi/\sqrt 2$\\
$\pi^+\pi^-$&$1/2$&$0$&$0$\\
$K^+K^-$&$s\bar s\to s\bar s:1$&$u\bar u\to u\bar
u:1$&$0$\\\hline\hline
\end{tabular}
\end{center}
\end{table}

\begin{table}
\begin{center}
\parbox{0.6\textwidth}{\caption{The harmonic parameters calculated with the mass $M$ and
the decay constant $f_M$ \cite{decay_constant} by the relation
$\beta=\sqrt \pi ({M\over 12})^{1/3}f_M^{2/3}$\label{harmonicpar}}}
\begin{tabular}{cccc}
\hline\hline Meson&Mass (GeV)&$f_M$&$\beta$ (GeV)\\\hline
$\pi$&$m_\pi$&$f_\pi=0.130$&0.103\\
$\eta_q$&$m_\pi$&$1.07f_\pi$&0.108\\
$\eta_s$&$\sqrt{2m_K^2-m_\pi^2}$&$1.34f_\pi$&0.213\\
$\rho$&$m_\rho$&$0.220$&0.258\\
$\omega$&$m_\omega$&$0.195$&0.240\\
$K$&$m_K$&$0.159$&0.180\\
$\phi$&$m_\phi$&$0.245$&0.305\\
$K^*$&$m_{K^*}$&$0.217$&0.270\\\hline\hline
\end{tabular}
\end{center}
\end{table}
For $X(1812)\to M_1M_2$ decays, the decay width includes the
contributions from the quark exchange and quark annihilation
mechanisms, respectively denoted by $\Gamma_{M_1M_2}^{\textrm{ex}}$
and $\Gamma_{M_1M_2}^{\textrm{an}}$ . Due to the non-zero flavor
factors, the quark exchange mechanism allows only three decays,
$\omega\phi,~\eta\eta$ and $\eta\eta'$ channels as shown in Table
\ref{flavorfactor}. The calculation of the decay width is
straightforward though tedious according to Eq. (\ref{cross}). The
numerical calculation yields the ratios:

\begin{eqnarray}
\Gamma_{\eta\eta'}^{\textrm{ex}}:\Gamma_{\eta\eta}^{\textrm{ex}}:\Gamma_{\omega\phi}^{\textrm{ex}}:\Gamma_{K^+K^-}^{\textrm{ex}}:\Gamma_{\rho^+\rho^-}^{\textrm{ex}}:\Gamma_{\omega\omega}^{\textrm{ex}}:\Gamma_{K^{*+}K^{*-}}^{\textrm{ex}}:\Gamma_{\pi^+\pi^-}^{\textrm{ex}}=\nonumber\\
9.3:4.3:1.0:0:0:0:0:0.
\end{eqnarray}
This indicates that the quark exchange mechanism enhances the decays
$X(1812)\to\eta\eta$ and $\eta\eta'$. On the other hand, the quark
annihilation mechanism makes no contribution to the decay
$X(1812)\to\omega\phi$. To determine the contributions from this
mechanism, we estimate the reduced ratios
$\tilde{\Gamma}_{M_1M_2}^{\textrm{an}}=\Gamma_{M_1M_2}^{\textrm{an}}/\Gamma_{\omega\phi}^{\textrm{ex}}$.
The numerical calculation yields:
\begin{eqnarray}
\tilde{\Gamma}_{\eta\eta'}^{\textrm{an}}:\tilde{\Gamma}_{\eta\eta}^{\textrm{an}}:\tilde{\Gamma}_{\omega\phi}^{\textrm{an}}:\tilde{\Gamma}_{K^+K^-}^{\textrm{an}}:\tilde{\Gamma}_{\rho^+\rho^-}^{\textrm{an}}:\tilde{\Gamma}_{\omega\omega}^{\textrm{an}}:\tilde{\Gamma}_{K^{*+}K^{*-}}^{\textrm{an}}:\tilde{\Gamma}_{\pi^+\pi^-}^{\textrm{an}}=\nonumber\\
0.39:0.04:0:0.44:0.41:0.005:0.001:1.9\times10^{-5}.
\end{eqnarray}
This indicates that the quark annihilation mechanism makes small
contributions to $X(1812)\to M_1M_2$, especially for the
$\pi^+\pi^-$ mode, which is highly suppressed in dynamics. If the
both decay mechanisms and the interference effects between them are
taken into consideration, the total decay widths for each mode are
calculated to be:
\begin{eqnarray}
\Gamma_{\eta\eta'}:\Gamma_{\eta\eta}:\Gamma_{\omega\phi}:\Gamma_{K^+K^-}:\Gamma_{\rho^+\rho^-}:\Gamma_{\omega\omega}:\Gamma_{K^{*+}K^{*-}}:\Gamma_{\pi^+\pi^-}=\nonumber\\
13.6:5.14:1:0.44:0.41:0.005:0.001:1.9\times10^{-5}
\end{eqnarray}

For the production of the $(\ksks)$ molecule in the
$\jp\to\gamma(\ksks)$, the production rate depends on the overlap of
the $K^*$ and $\bar K^*$ wave functions in momentum space. They are
phenomenologically described by ground harmonic oscillator bases
with a parameter $\beta$, which is related to the root mean square
(rms) radius $\langle r^2\rangle$ by the relation of
$\beta=\sqrt{{3\over 2 \la r^2\ra}}$. The model study using a
nonrelativistic Hamiltonian with pairwise effective interactions
shows that the rms becomes large when the binding energy of
two-meson molecule decreases \cite{wongcy}. With the binding energy
ranging from $100$ MeV to a few Mev, the rms runs from 0.3fm to $2$
fm. Near the threshold of two bound mesons, the rms of the molecule
is about 1-2 fm. In our calculation, we naively set the $\beta$
parameter within the range $0.5\sim 2.0$ fm. From Eq.
(\ref{prod1}-\ref{prod2}) the numerical calculation yields the
ratios of the decay width for the $\jp\to \gamma\ksks$ and
$\gamma(\ksks)$, which are given in Table \ref{prd}. The results
show that the strong binding of the $(K^{*}\bar K^{*})$ molecule
favors its production from $\jp$ decays.

\begin{table}
\begin{center}
\parbox{0.6\textwidth}{\caption{The production rate of $\Gamma_{\gamma
(\ksks)}/\Gamma_{\gamma K^*\bar K^*}$ in terms of the rms of the
$(\ksks)$ molecule. \label{prd}}}
\begin{tabular}{cccccc}
\hline\hline
 $\sqrt{\la r^2\ra}$fm&0.5&0.8&1&1.5&2.0\\\hline
$\Gamma_{\gamma (\ksks)}/\Gamma_{\gamma K^*\bar
K^*}$&9.54&3.79&2.13&0.66&0.28\\
\hline\hline
\end{tabular}
\end{center}
\end{table}
\section{Discussion and Summary  }
Based on the scenario of $X(1812)$ to be a $(\ksks)$ molecule, the
decay width and the production rate are evaluated. It is found that
the quark exchange mechanism plays an important role in the
understanding of the decay $X(1812)\to\omega\phi$, and we find that
this decay mechanism enhances the decay widths for $\eta\eta$ and
$\eta\eta'$ modes. On the other hand, the quark annihilation
mechanism makes a small contributions to $X(1812)\to PP$ and $VV$
decays, and the decay $X(1812)\to\pi\pi$ is highly suppressed in
dynamics. These results suggest that if $X(1812)$ is a $(\ksks)$
molecule, it should be found in the decays $\jp\to\gamma
X,~X\to\eta\eta'$ and $\eta\eta$ except for the observed mode
$X\to\omega\phi$. Unfortunately, due to the low statistics of the
present data sample, no information on $\jp\to\gamma\eta\eta'$ and
$\gamma\eta\eta$ decays are available in PDG table.

\begin{figure}[htbp]
 \vspace*{0cm}
\begin{center}
\hspace*{-.cm} \epsfysize=13cm \epsffile{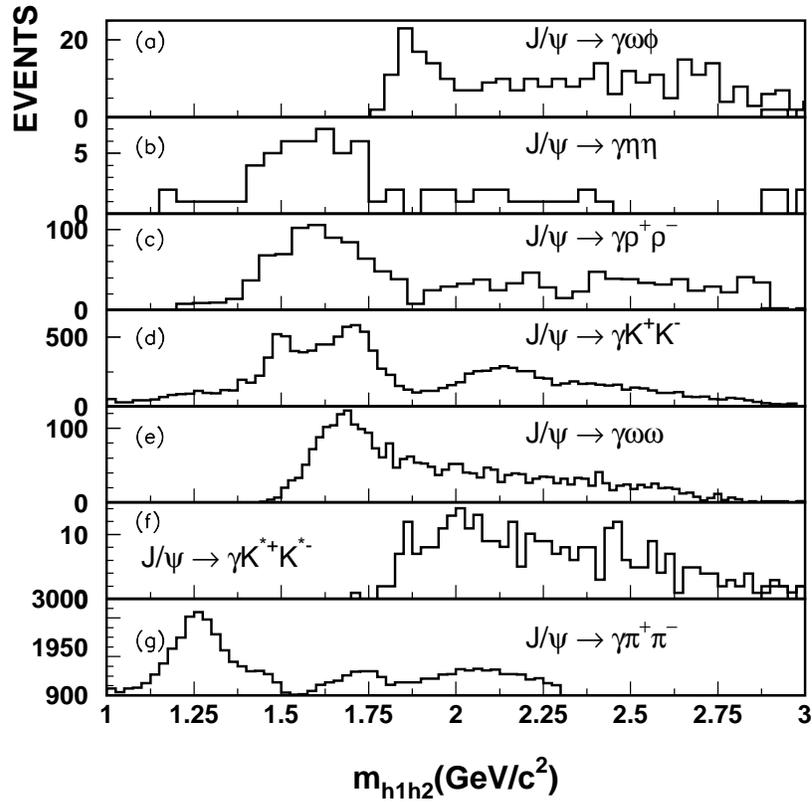}
\parbox{0.7\textwidth}{\vspace*{-1cm}\caption{The experimental measurement of mass distributions
$m_{h_1h_2}$ for radiative decays of $\jp\to\gamma h_1h_2$, (a)
$\jp\to\gamma \omega\phi$\cite{bes_gop}, (b) $\jp\to\gamma \eta\eta$
\cite{getaeta}, (c) $\jp\to\gamma \rho^+\rho^-$ \cite{markiii_grr},
(d) $\jp\to\gamma K^+K^-$ \cite{bes_gkk}, (e) $\jp\to\gamma
\omega\omega$\cite{bes_goo}, (f) $\jp\to\gamma K^{*+}K^{*-}$
\cite{bes_gksks}, (g) $\jp\to\gamma \pi^+\pi^-$ \cite{bes_gpipi}.
\label{exp_all}}}
\end{center}
\end{figure}

Figure \ref{exp_all} shows experimental information on the invariant
mass distribution of two mesons in
$\jp\to\gamma\omega\phi,~\gamma\eta\eta,~\gamma\rho^+\rho^-,~\gamma
K^+K^-,~\gamma\omega\omega,~\gamma K^{*+}K^{*-}$ and
$\gamma\pi^+\pi^-$. In $X(1812)$ mass region as seen in the
$\jp\to\gamma\omega\phi$, no signal evidence is observed in the
other decays. As for the decay $\jp\to \gamma \eta\eta$, it is
firstly reported by the collaboration of Crystal Ball detector
twenty years ago with very low statistics, but after that no
confirmation was made by other collaborations. Our predictions
deserve to be tested in this decay, together with the
$\jp\to\gamma\eta\eta'$. With a large data sample accumulation and
the improvement in the detector performance, especially for the
photon identification for BESIII and CLEOc detector, search for
$X(1812)$ signals in these two channels are possibly achieved. In
the $\jp\to\gamma\rho^+\rho^-$, the partial wave analysis (PWA)
shows that the contribution from scalar near the mass of 1690 MeV
dominates the $m_{\rho\rho}$ mass distribution. In the
$\jp\to\gamma\pi^+\pi^-$, the PWA shows that significant resonance
with mass 1270 Mev favors the $J^{PC}=2^{++}$, and small scalar
components with mass equal to 1466 MeV and 1765 MeV are also
reported. Recently, the BESII collaboration reported the PWA
performed on the $\jp\to\gamma\omega\omega$. The results show that
the contributions in the mass distribution of $m_{\omega\omega}$
below $2.0$ GeV are predominantly from pseudoscalars, and only with
small contributions from $f_0(1710),~f_2(1640)$ and $f_2(1910)$. The
PWA of the decay $\jp\to\gamma K^+K^-$ is also reported by the BESII
collaboration, the dominant contributions are from the scalar
$f_0(1710)$ and the tensor $f_2'(1525)$. The PWA on the decay
$\jp\to\gamma \ksks$ is reported by BESI collaboration with 7.8
million $\jp$ data, the results showed that a broad $0^{-+}$
resonance with mass $M=1800$ MeV is observed, no significant
$0^{++},~1^{++}$ or $4^{++}$ signal is found. In short, no evidence
for $X(1812)$ is currently observed in the hadron mass distributions
of the radiative decays: $\jp\to\gamma\rho^+\rho^-,~\gamma
K^+K^-,~\gamma\omega\omega,~\gamma K^{*+}K^{*-}$ and
$\gamma\pi^+\pi^-$, from which it seems to be consistent with the
our calculation based on the $(\ksks)$ molecular picture. The
further test of this scenario is expected to search for the
$X(1812)$ in $\jp\to\gamma\eta\eta'$ and $\gamma\eta\eta$ decays in
the future.

If we naively assume that the $(K^{*+}K^{*-})$ molecule dominantly
decays into $\eta\eta',~\eta\eta,\omega\phi,~K\bar K$ and
$\rho\rho$, then one obtains $Br(X\to\omega\phi)\simeq 4.60\%$ from
our calculations. From the measurement of $Br(\jp\to\gamma
X)Br(X\to\omega\phi)=(2.61\pm0.71)\times 10^{-4}$, one has the
production rate $Br(\jp\to\gamma X)=(5.67\pm 0.07)\times 10^{-3}$.
Combined with the PDG value, one has $\Gamma(\jp\to\gamma
X)/\Gamma(\jp\to\gamma K^{*+}K^{*-})=2.83\pm0.92$. Our calculation
yields $\Gamma(\jp\to\gamma X)/\Gamma(\jp\to\gamma
K^{*+}K^{*-})=2.13^{+7.41}_{-1.85}$, where the central value
corresponds to the $\sqrt{\la r^2\ra}=1.0$ fm, and the uncertainty
corresponds to the range $0.5\textrm{~fm}\le \sqrt{\la r^2\ra} \le
2.0\textrm{~fm}$.

It is also important to look for decay rates of $X(1812)\to PP$ and
$VV$ in experiment for understanding its nature. Theoretically,
these decays have been evaluated numerically by many authors using
different models to describe the $X(1812)$ structure. For example,
in four-quark picture \cite{fq}, it is found that
$X(1812)\to\omega\omega,~\ksks$ are the two dominant decay channels
and $X(1812)\to K\bar K,~\eta\eta,~\eta\eta'$ are suppressed. While
in glueball picture \cite{gl1}, the dominant decays are
$X(1812)\to\rho\rho$ and $\omega\omega$, and decays into
$\omega\phi$ and $\ksks$ are highly suppressed. In the
quarkonia-glueball-hybrid mixing scheme \cite{hy2}, it turns out
that the branching fractions for $X(1812)\to K\bar K$ and $\eta\eta$
are about 30\%, which adds up to about 70\% of the total $X(1812)$
decay width. However, in our $(\ksks)$ molecule picture, the
dominate decays are $X(1812)\to \eta\eta'$ and $\eta\eta$.

In summary, we evaluate the decay widths of
$X(1812)\to\eta\eta',~\eta\eta,~\omega\phi,~K^+K^-,~\rho^+\rho^-,~\omega\omega,~K^{*+}K^{*-}$
and $\pi^+\pi^-$ based on the assumption of the $X(1812)$ as a
candidate of $(\ksks)$ molecule. It turns out that the quark
exchange mechanism plays an important role in the understanding of
the large decay with for the $X(1812)\to \omega\phi$. We  also find
that the decay widths for $X(1812)\to\eta\eta'$ and $\eta\eta$
decays are enhanced by the quark exchange mechanism. These channels
are suggested to be the tools to test the molecular scenario in
experiment. The branching fraction of $Br(X\to\omega\phi)$ is
evaluated to be about $4.60\%$. Searches for additional evidence
about the $X(1812)$ in $\jp$ radiative decays are reviewed. In the
molecular scenario, the $X(1812)$ production rate is also evaluated
to be $\Gamma(\jp\to\gamma X)/\Gamma(\jp\to\gamma
K^{*+}K^{*-})=2.13^{+7.41}_{-1.85}$, which is close to the measured
value $2.83\pm0.92$.

\vspace{1cm}
{\large Acknowledgements}\\
The work is partly supported by the National Natural Science
Foundation of China under Grant Nos. 10575083, 10435080, 10375074
and 10491303.

\appendix
\section{The amplitude for $gg\to K^*\bar K^*$}
We start with the process $g(\epsilon_1)g(\epsilon_2)\to
q(p_1,s_1)\bar q(q_1,\bar s_1)q(p_2,s_2)\bar q(q_2,\bar s_2)\to
K^*({\bf P_1})\bar K^*({\bf P_2})$ as shown in Fig. \ref{figprd},
where $\epsilon_i~(i=1,2)$ denotes the polarization vector of the
gluon; the momentum and the spin of the quark (antiquark) $q~(\bar
q)$ are denoted by $p_i~(q_i)$ and $s_i~(\bar s_i)$, respectively;
 the outgoing momentum for $K^*~(\bar K^*)$ is denoted by ${\bf
P_1}~({\bf P_2})$. The amplitude with Lorentz indexes $\mu_1$ and
$\mu_2$ is defined as:
\begin{eqnarray}\label{ggamp}
A_{gg\to\ksks}^{\mu_1\mu_2}&=&\int{1\over
\sqrt{2p_1^02q_1^02p_2^02p_2^0}}{d^3p_1\over (2\pi)^{3}}{d^3q_1\over
(2\pi)^{3}}{d^3p_2\over (2\pi)^{3}}{d^3q_2\over (2\pi)^{3}}\nonumber\\
&\times&\sum_{\textrm{all spin indexes}}\bar
u(s_1,p_1)\gamma^{\mu_1}v(\bar
s_2,q_2)u(s_2,p_2)\gamma^{\mu_2}v(\bar s_1,q_1)\nonumber \\
&\times&\cg{\half s_1}{\half\bar s_1}{S_1S_{1z}}\cg{\half
s_2}{\half\bar
s_2}{S_2S_{2z}}\cg{L_1M_1}{S_1S_{1z}}{J_1J_{1z}}\cg{L_2M_2}{S_2S_{2z}}{J_2J_{2z}}\nonumber\\
&\times&\Psi_{K^*L_1S_1}({\bf p_1-q_1})\Psi_{\bar K^*L_2S_2}({\bf
p_2-q_2})\nonumber\\
&\times&(2\pi)^3\delta({\bf P_1-p_1-q_1})(2\pi)^3\delta({\bf
P_2-p_2-q_2}),
\end{eqnarray}
where $\Psi_{K^*L_1S_1}~(\Psi_{\bar K^*L_2S_2})$ denotes the wave
function for $K^*~(\bar K^*)$ with the orbit and spin
angular-momentum quantum number $L_1~(L_2)$ and $S_1~(S_2)$,
respectively. For the vector meson $K^*$, it is assigned as the
$^3S_1$ state in $q\bar q$ quark model. So we have $L_i=0$ and
$S_i=J_{iz}=1~(i=1,2)$.

The above amplitude can be simplified with the help of the spin
projection operator $P_{S_1S_{1z};S_2S_{2z}}^{\mu_1\mu_2}$, which is
defined as:
\begin{eqnarray}\label{proj}
P_{S_1S_{1z};S_2S_{2z}}^{\mu_1\mu_2}&\equiv&\sum_{s_1\bar s_1
s_2\bar s_2}u(s_1,p_1)\gamma^{\mu_1}v(\bar
s_2,q_2)u(s_2,p_2)\gamma^{\mu_2}v(\bar s_1,q_1)\nonumber \\
&\times&\cg{\half s_1}{\half\bar s_1}{S_1S_{1z}}\cg{\half
s_2}{\half\bar s_2}{S_2S_{2z}}\nonumber \\
&=&\textrm{Tr}\left [\sum_{s_1\bar s_1}v(\bar
s_1,q_1)u(s_1,p_1)\gamma^{\mu_1}\cg{\half s_1}{\half\bar
s_1}{S_1S_{1z}}\right.\nonumber \\
&\times&\left.\sum_{s_2\bar s_2}v(\bar
s_2,q_2)u(s_2,p_2)\gamma^{\mu_2}\cg{\half s_2}{\half\bar
s_2}{S_2S_{2z}}\right ]\nonumber\\
&\equiv& \textrm{Tr}[P_{1S_{1z}}P_{1S_{2z}}],
\end{eqnarray}
where $P_{1S_{1z}}$ and $P_{1S_{2z}}$ are the spin projection
operators for the spin-1 particles. As given in \cite{GKPR}, they
are explicitly expressed with the spin polarization vectors
$\epsilon_{K^*}$ as:
\begin{eqnarray}
P_{1S_z}^\mu(p_i,q_i)&\equiv&\sum_{s_i\bar s_i}v(\bar
s_i,q_i)u(s_i,p_i)\gamma^{\mu}\cg{\half s_i}{\half\bar
s_i}{1S_{z}}\nonumber \\
&=&{1\over \sqrt{(q_i^0+m_s)(p_i^0+m_u)}}(-\sla
q_i+m_s)\epsilon_{K^*}(S_z)(\sla p+m_u).
\end{eqnarray}
To simplify the integral in Eq. (\ref{ggamp}), one method to make an
approximation is the nonrelativistic assumption for valence quarks,
ie. ${\bf p_i-q_i}=2{\bf q}\rightarrow 0$. Then the integral of the
amplitude can be evaluated at the origin with the substitution of
the momentum $p_i={m_uP_i\over m_u+m_s}$ and $q_i={m_sP_i\over
m_u+m_s}$, ie.
\begin{eqnarray}\label{appro}
\int{d^3p_1\over (2\pi)^{3}}{d^3q_1\over
(2\pi)^{3}}\Psi_{K^*0S}({\bf p_1-q_1})(2\pi)^3\delta({\bf
P_1-p_1-q_1})\approx {R_{K^*0}(0)\over \sqrt{4\pi}},
\end{eqnarray}
where $R_{K^*0}(0)$ is the radius wavefunction for the $K^*$ at
origin. With the help of equations (\ref{proj}) to (\ref{appro}),
the amplitude for $gg\to \ksks$ can be simplified as:
\begin{eqnarray}\label{}
 A_{gg\to K^*\bar K^*}^{\mu_1,\mu_2}\epsilon_{1\mu_1}\epsilon_{2\mu_2}&=&{1\over 2!}{1\over 4\sqrt{p_1^0p_2^0q_1^0q_2^0}}
 {1\over
 \sqrt{(q_1^0+m_s)(q_2^0+m_s)(p_1^0+m_u)(p_2^0+m_u)}}{R_{K^*}^2(0)\over 4\pi}\nonumber \\
&\times& \textrm{Tr}[(-\sla q_1+m_s)\sla \epsilon_{K^*}(\sla
p_1+m_u)\sla\epsilon_2(-\sla q_2+m_s)\epsilon_{\bar K^*}(\sla
p_2+m_u)\sla \epsilon_1\nonumber \\
&+&(\epon\leftrightarrow \eptw)],
\end{eqnarray}
where the factor of $1/2!$ comes from the contribution of the cross
term for the two gluons of identical particles.

\end{document}